\documentclass[final,3p,times,sort&compress]{elsarticle}

\usepackage{amsmath}
\usepackage{amssymb}
\usepackage{epsfig}

\journal{Physics Letters B}

\begin{document}

\begin{frontmatter}



\title{QCD predictions for the azimuthal asymmetry in charm leptoproduction for the COMPASS kinematics}


\author[dubna]{A.V.~Efremov}
\ead{efremov@theor.jinr.ru}
\author[yerphi]{N.Ya.~Ivanov\corref{cor1}}
\cortext[cor1]{Corresponding author.}
\ead{nikiv@yerphi.am}
\author[dubna]{O.V.~Teryaev}
\ead{teryaev@theor.jinr.ru}

\address[dubna]{Bogoliubov Laboratory of Theoretical Physics, JINR, 141980 Dubna, Russia}
\address[yerphi]{Yerevan Physics Institute, Alikhanian Br.~2, 0036 Yerevan, Armenia}

\begin{abstract}
We present the QCD predictions for the azimuthal $\cos 2\varphi$ asymmetry in charm leptoproduction for the kinematics of the COMPASS experiment at CERN. The asymmetry is predicted to be large, about 15\%. The radiative corrections to the QCD predictions for the $\cos 2\varphi$ distribution are estimated to be small, less than 10\%. Our calculations show that the azimuthal asymmetry in charm production is well defined in pQCD: it is stable both perturbatively and parametrically, and practically insensitive to theoretical uncertainties in the input parameters. We analyze the nonperturbative contributions to the $\cos 2\varphi$ distribution due to the gluon transverse motion in the target and the $c$-quark fragmentation. Because of the $c$-quark low mass, the nonperturbative contributions are expected to be sizable, about (30--40)\%. We conclude that extraction of the azimuthal asymmetries from available COMPASS data will provide valuable information about the transverse momentum dependent distribution of the gluon in the proton and the $c$-quark hadronization mechanism. Finally, we discuss the $\cos 2\varphi$ asymmetry as a probe of the gluonic analogue of the Boer-Mulders function, $h_{1}^{\perp g}$, describing the linear polarization of gluons inside unpolarized proton.

\end{abstract}

\begin{keyword}
QCD \sep Charm Leptoproduction \sep Azimuthal Asymmetry \sep COMPASS
\PACS 12.38.Bx \sep 13.60.Hb \sep 13.88.+e

\end{keyword}

\end{frontmatter}


\section{Introduction and notation} 
\label{1.0}
Perturbatively stable observables in charm production are of special interest in investigation of the role of nonperturbative effects at a GeV scale. Measurements of the quantities which are stable under radiative corrections could provide direct access to nonperturbative contributions. Nontrivial examples of perturbatively stable observables were proposed in Refs.~\cite{we1,we2,we4,we7,we8,Almeida-S-V}, where, in particular, the azimuthal $\cos 2\varphi$ asymmetry (AA) in heavy-quark leptoproduction was analyzed. It was shown in Refs.~\cite{we1,we2,we4} that, contrary to the production cross sections, the azimuthal asymmetry in charm photo- and leptoproduction is quantitatively well defined in pQCD: the contribution of the dominant photon-gluon fusion mechanism to the AA is stable, both parametrically and perturbatively. In this paper, we show that the AA in charm leptoproduction is sensitive to nonperturbative contributions:  the gluon transverse motion in the target and $c$-quark fragmentation. For this reason, measurements of the azimuthal asymmetry will provide valuable information about the transverse momentum dependent (TMD) distribution of the gluon in the proton and the $c$-quark hadronization mechanism.

In recent article \cite{Compass_2012}, the COMPASS Collaboration has presented the semi-inclusive differential distributions of charmed mesons produced by 160 GeV muons in DIS. 
In particular, semi-inclusive spectra of $D^{*\pm}$ mesons are given as a function of their  lab system energy $E$, transverse momentum $p_{\perp}$, energy fraction $z$, and virtual photon energy $\nu$. The COMPASS data are concentrated in the kinematic range defined by 0.003 GeV$^2$~$< Q^2 <$~10 GeV$^2$, $3\cdot 10^{-5} < x < 0.1$ and 20~GeV $< E <$~80 GeV.
 
\begin{figure*}
\begin{center}
\mbox{\epsfig{file=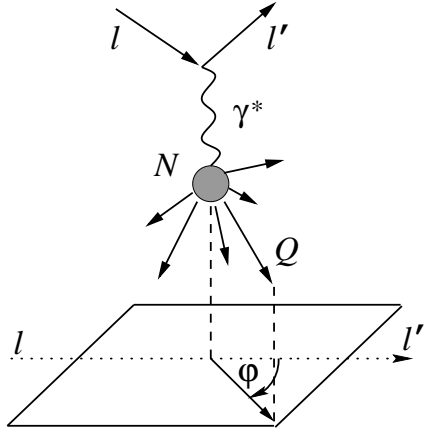,width=200pt}}
\caption{\label{Fg.1}\small Definition of the azimuthal angle $\varphi$ in the nucleon rest frame.}
\end{center}
\end{figure*}

In the present paper, we provide the QCD predictions for the azimuthal $\cos 2\varphi$ asymmetry in charm leptoproduction, 
\begin{equation}
l(\ell )+N(P)\rightarrow l^{\prime}(\ell -q)+Q(p_{Q})+X[\bar{Q}](p_{X}), \label{1}
\end{equation}
for the COMPASS kinematics. Neglecting the contribution of $Z$-boson exchange, the azimuth-dependent cross section of the reaction (\ref{1}) can be written as
\begin{eqnarray}
\frac{{\rm d}^{5}\sigma_{lN}}{{\rm d}x\,{\rm d}Q^{2}{\rm d}T_{1}{\rm d}U_{1}{\rm d}\varphi}=\frac{\alpha_{em}}{(2\pi)^{2}}\frac{1}{x\,Q^{2}}\frac{y^2}{1-\varepsilon}\Biggl[\frac{{\rm d}^{2}\sigma_{T}}{{\rm d}T_{1}{\rm d}U_{1}}\left(x,Q^{2},T_{1},U_{1}\right)+ \varepsilon\frac{{\rm d}^{2}\sigma_{L}}{{\rm d}T_{1}{\rm d}U_{1}}\left(x,Q^{2},T_{1},U_{1}\right) \Biggr. \label{2} \\
+\Biggl.\varepsilon\frac{{\rm d}^{2}\sigma_{A}}{{\rm d}T_{1}{\rm d}U_{1}}\left(x,Q^{2},T_{1},U_{1}\right)\cos 2\varphi+2\sqrt{\varepsilon(1+\varepsilon)}\frac{{\rm d}^{2}\sigma_{I}}{{\rm d}T_{1}{\rm d}U_{1}}\left(x,Q^{2},T_{1},U_{1}\right)\cos \varphi\Biggr], \nonumber
\end{eqnarray}
where $\alpha_{\mathrm{em}}$ is Sommerfeld's fine-structure constant, the quantity $\varepsilon$ measures the degree of the longitudinal polarization of the virtual photon in the Breit frame \cite{dombey}, $\varepsilon=\frac{2(1-y)}{(1+(1-y)^2)}$, and the kinematic variables are defined by
\begin{align}
\bar{S}&=2\left( \ell\cdot P\right), & y&=\frac{q\cdot P}{\ell\cdot P },& T_{1}&=\left(P-p_{Q}\right)^{2}-m^{2},  \notag \\
Q^{2}&=-q^{2}, & x&=\frac{Q^{2}}{2q\cdot P},& U_{1}&=\left( q-p_{Q}\right)^{2}-m^{2}.  \label{3}
\end{align}
In the nucleon rest frame, the azimuth $\varphi$ is the angle between the lepton scattering plane and the heavy quark production plane, defined by the exchanged photon and the detected quark $Q$ (see Fig.~\ref{Fg.1}). 
The covariant definition of $\varphi $ is
\begin{align}
\cos \varphi &=\frac{r\cdot n}{\sqrt{-r^{2}}\sqrt{-n^{2}}},&
\sin \varphi &=\frac{Q^{2}\sqrt{1/x^{2}+4m_{N}^{2}/Q^{2}}}{2\sqrt{-r^{2}}\sqrt{-n^{2}}}~n\cdot \ell, \notag \\
r^{\mu } &=\varepsilon ^{\mu \nu \alpha \beta }P_{\nu }q_{\alpha }\ell _{\beta },& n^{\mu }&=\varepsilon ^{\mu \nu \alpha \beta }q_{\nu }P_{\alpha }p_{Q\beta }.\label{4}
\end{align}
In Eqs.~(\ref{3}) and (\ref{4}), $m$ and $m_{N}$ are the masses of the heavy quark and the  target, respectively. In terms of the above definitions, the variables under consideration are:
\begin{align}
E&=\frac{m_{N}^{2}-T_{1}}{2m_{N}}, \qquad \quad \nu=\frac{y\bar{S}-m_{N}^{2}}{2m_{N}}, \qquad \quad z=\frac{E}{\nu}, \notag \\ 
p^{2}_{\perp}&=\frac{T_{1}U_{1}}{y\bar{S}}+\frac{T_{1}Q^{2}}{y^{2}\bar{S}^{2}}\left(y\bar{S}+T_{1}\right)-m^{2}. \label{5} 
\end{align} 

In Eq.~(\ref{2}), ${\rm d}^{2}\sigma_{T}\,({\rm d}^{2}\sigma_{L})$ is the usual $\gamma ^{*}N$ cross section describing heavy quark production by a transverse (longitudinal) virtual photon. The cross section ${\rm d}^{2}\sigma_{A}\,({\rm d}^{2}\sigma_{I})$ originates from interference between transverse (longitudinal and transverse) components of the photon and is responsible for the $\cos2\varphi$ ($\cos\varphi$) asymmetry.

\section{Parton-level cross sections} 
\label{2.0}
\subsection{LO predictions}
\label{2.1}

Within the dominant photon-gluon fusion mechanism, the hadron-level cross sections, $\frac{{\rm d}^{2}\sigma_{k}}{{\rm d}T_{1}{\rm d}U_{1}}\left(x,Q^{2},T_{1},U_{1}\right)~~(k=T,L,A,I)$, are related to the partonic ones, $\frac{{\rm d}^{2}\hat{\sigma}_{k}}{{\rm d}t_{1}{\rm d}u_{1}}\left(\hat{x},Q^{2},t_{1},u_{1}\right)$, as follows:
\begin{equation}
\frac{{\rm d}^{2}\sigma_{k}}{{\rm d}T_{1}{\rm d}U_{1}}\left(x,Q^{2},T_{1},U_{1}\right)=\int^{1}_{\zeta^{-}}{\rm d}\zeta\,\zeta\, g(\zeta,\mu_F)\frac{{\rm d}^{2}\hat{\sigma}_{k}}{{\rm d}t_{1}{\rm d}u_{1}}\left(x\left/\right.\zeta, Q^{2},\zeta T_{1},U_{1}\right) \qquad (k=T,L,A,I), \label{6}
\end{equation}
where $\zeta^{-}=-\frac{xU_{1}}{Q^2+xT_{1}}$, $g(\zeta,\mu_F)$ describes gluon density in the proton evaluated at a factorization scale $\mu_F$, and partonic invariants in the single-particle inclusive (1PI) kinematics are:
\begin{equation}
\hat{x}=x\left/\right.\zeta,\qquad \qquad t_{1}=\zeta T_{1},\qquad \qquad u_{1}=U_{1}. \label{7}
\end{equation}

At leading order, ${\cal O}(\alpha _{em}\alpha_{s})$, the parton-level cross sections have the form:
\begin{equation}
\frac{{\rm d}^{2}\hat{\sigma}_{k}^{\rm Born}}{{\rm d}t_{1}{\rm d}u_{1}}(\hat{x},Q^{2} ,t_{1},u_{1})=\pi e_{Q}^{2}\alpha_{em}\alpha_{s}\frac{\hat{x}^{2}}{Q^{4}} B_{k}(\hat{x},Q^{2},t_{1},u_{1})\,\delta(Q^2/\hat{x}+t_{1}+u_{1}),\label{8}
\end{equation}
where
\begin{eqnarray}
B_{T}(\hat{x},Q^{2},t_{1},u_{1}) &=&\frac{t_{1}}{u_{1}}+\frac{u_{1}}{t_{1}%
}+4\left(1-\hat{x}-\frac{m^{2}Q^2}{\hat{x}\, t_{1}u_{1}}\right)
\left( \frac{Q^2(m^{2}-Q^{2}/2)}{\hat{x}\, t_{1}u_{1}}+\hat{x}\right),  \nonumber \\
B_{L}(\hat{x},Q^{2},t_{1},u_{1}) &=&8\hat{x}\left(1-\hat{x}-\frac{m^{2}Q^2}{\hat{x}\, t_{1}u_{1}}\right), 
\nonumber \\
B_{A}(\hat{x},Q^{2},t_{1},u_{1}) &=&4\left(1-\hat{x}-\frac{m^{2}Q^2}{\hat{x}\, t_{1}u_{1}}\right) \left(\frac{m^{2}Q^2}{\hat{x}\,t_{1}u_{1}}+\hat{x}\right),  \label{9} \\
B_{I}(\hat{x},Q^{2},t_{1},u_{1}) &=&4\sqrt{Q^{2}}\left(\hat{x}(1-\hat{x})\frac{t_{1}u_{1}}{Q^2}-m^{2}\right) ^{1/2}\frac{u_{1}-t_{1}}{t_{1}u_{1}}\left(1-2\hat{x}-\frac{2m^{2}Q^2}{\hat{x}\, t_{1}u_{1}}\right).  \nonumber
\end{eqnarray}

\subsection{NLO corrections}
\label{2.2}
The exact NLO, ${\cal O}(\alpha _{em}\alpha_{s}^2)$, contributions to the $\hat{\sigma}_{T}(\hat{x},Q^{2})$ and $\hat{\sigma}_{L}(\hat{x},Q^{2})$ cross sections have been calculated  in Ref.~\cite{LRSN}. They are presently available in the form of fast computer codes, see e.g. Ref.~\cite{Blumlein}.

The exact NLO predictions for the azimuth dependent cross sections $\hat{\sigma}_{A}(\hat{x},Q^{2})$ and $\hat{\sigma}_{I}(\hat{x},Q^{2})$ are presently unavailable. They, however, can be estimated within the soft-gluon (or threshold) approximation. These soft-gluon corrections to $\hat{\sigma}_{A}(\hat{x},Q^{2})$ have been studied to next-to-leading logarithmic (NLL) accuracy in Ref.~\cite{we4}. In particular, the NLO NLL corrections have been calculated to the hadron-level azimuthal $\cos 2\varphi$ asymmetry, $A(x,Q^2)$, defined as
\begin{equation}
A(x,Q^2)=\frac{\sigma_{A}}{\sigma_{2}}(x,Q^{2}),  \label{10}
\end{equation}
where $\sigma_{2}(x,Q^2)=\sigma_{T}(x,Q^2)+\sigma_{L}(x,Q^2)$.

\begin{figure*}
\begin{center}
\mbox{\epsfig{file=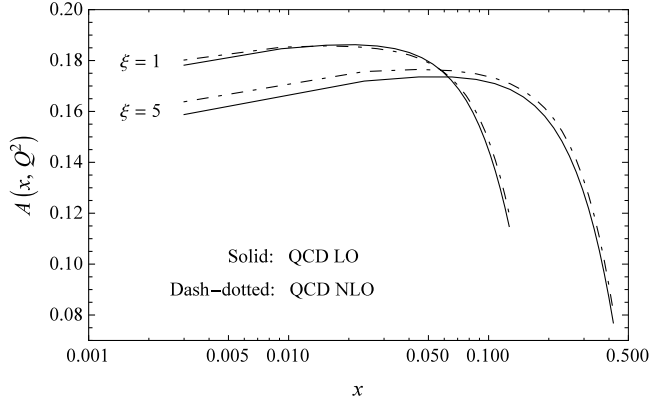,width=240pt}}
\caption{\label{Fg.2}\small LO (solid lines) and NLO (dash-dotted lines) soft-gluon predictions for the $x$ dependence of the azimuthal $\cos2\varphi$ asymmetry,
$A(x,Q^{2})=\sigma_{A}/\sigma_{2}$, in charm leptoproduction at $\xi$=1 and 5, $\xi=Q^2/m^2$.}
\end{center}
\end{figure*}

Our results for the $x$ distribution of the asymmetry $A(x,Q^{2})$ in charm leptoproduction at fixed values of $\xi=\frac{Q^2}{m^2}$ are presented in Fig.~\ref{Fg.2}. One
can see that, in the kinematics of the COMPASS experiment,\footnote{In the case of charm production, values $\xi$=1 and $\xi$=5 correspond to $Q^2\approx$\,1.6 GeV$^2$ and $Q^2\approx$\,7.8 GeV$^2$, respectively.} the soft-gluon corrections to the production cross sections affect the Born predictions for $A(x,Q^2)$ at NLO very little, by a few percent only. For this reason, we neglect the radiative corrections to the azimuthal asymmetry in our further analysis.

\subsection{Nonperturbative contributions} 
\label{2.3}
Let us discuss how the photon-gluon fusion predictions for the azimuthal asymmetry are affected by nonperturbative contributions due to the gluon transverse motion in the target and $c$-quark fragmentation. Because of the $c$-quark low  mass, these contributions are especially important in description of the charmed particles production.

To introduce the $k_{T}$ degrees of freedom for initial gluon, $\vec{k}_{g}\simeq \zeta\vec{p}+\vec{k}_{T}$, one extends the integral over the parton distribution function in Eq.~(\ref{6}) to $k_{T}$-space,
\begin{equation}  \label{11}
{\rm d}\zeta\,g(\zeta,\mu _{F})\rightarrow {\rm d}\zeta\,{\rm d}^{2}k_{T}f_{1}^{g}\big(\zeta,\vec{k}^{2}_{T},\mu_{F}\big),
\end{equation}
where $f_{1}^{g}\big(\zeta,\vec{k}^{2}_{T},\mu_{F}\big)\simeq f\big(\vec{k}^{2}_{T}\big)g(\zeta,\mu_{F})$. The transverse momentum distribution, $f\big(\vec{k}^{2}_{T}\big)$, is usually taken to be a Gaussian:\footnote{The Gauss Ansatz for the TMD PDFs is used in many approaches, see e.g. Refs.~\cite{Efremov-TMD-1,Efremov-TMD,Mulders-TMD}. Note also that the Gaussian shape is supported by data \cite{Data-TMD}. For more details, see Ref.~\cite{Collins-TMD} and references therein.}
\begin{equation}  \label{12}
f\big(\vec{k}^{2}_{T}\big)=\frac{{\rm {e}}^{-\vec{k}_{T}^{2}/\langle k_{T}^{2}\rangle}}{\pi \langle k_{T}^{2}\rangle}.
\end{equation}
In our study, the analytic treatment of $k_{T}$ effects is used. According to Ref.~\cite{kT}, the $k_{T}$-smeared differential cross section of the process (\ref{1}) is a two-dimensional convolution:
\begin{equation}  \label{13}
\frac{{\rm d}^{2}\sigma_{k}^{\rm {kick}}}{{\rm d}^{2}p_{Q\perp}}\left(\vec{p}_{Q\perp}\right) =\int \text{d}^{2}k_{T} \frac{{\rm {e}}^{-\vec{k}_{T}^{2}/\langle k_{T}^{2}\rangle }}{\pi \langle k_{T}^{2}\rangle }\frac{\text{d}^{2}\sigma_{k}}{{\rm d}^{2}p_{Q\perp}}\Big(\vec{p}_{Q\perp}-\frac{1}{2}\vec{k}_{T}\Big) \qquad (k=T,L,A,I).
\end{equation}
The factor $\frac{1}{2}$ in front of $\vec{k}_{T}$ in the r.h.s. of Eq.~(\ref {13}) reflects the fact that the heavy quark carries away about one half of the initial energy in the reaction (\ref{1}). In numerical estimates, we use the value $\langle k_{T}^{2}\rangle=0.7$ GeV$^2$.

Hadronization effects in heavy flavor production are usually modeled with the help of the Peterson fragmentation function \cite{Peterson},
\begin{equation} \label{14}
D\,(\zeta)=\frac{a_{\varepsilon}}{\zeta\left[1-1/\zeta-\varepsilon/(1-\zeta)\right]^{2}},
\end{equation}
where $a_{\varepsilon }$ is a normalization factor and $\varepsilon_{D}=0.03$ in the case of a $D$-meson production. The hadron-level differential distribution has the following form: 
\begin{equation} \label{15}
\frac{{\rm d}^{3}\sigma _{k}^{D}}{{\rm d}^{3}p}\left(\vec{p}\right) =\int{\rm d}\zeta\,{\rm d}^{3}p_{Q}D\,(\zeta)\frac{{\rm d}^{3}\sigma _{k}}{{\rm d}^{3}p_{Q}}\left(\vec{p}_{Q}\right)\delta^{3}\left(\vec{p}-\zeta\vec{p}_{Q}\right) \qquad (k=T,L,A,I),
\end{equation}
where $\sigma _{lN}^{D}$ is the cross section for the production of the charmed meson $D$  with momentum $\vec{p}$, and $\sigma_{lN}$ is the cross section for the production of the $c$-quark with momentum $\vec{p}_Q$.

\section{Hadron-level results}
\label{3.0}
\subsection{Azimuthal asymmetries}
\label{3.1}
\begin{figure}[t]
\begin{center}
\begin{tabular}{cc}
\mbox{\epsfig{file=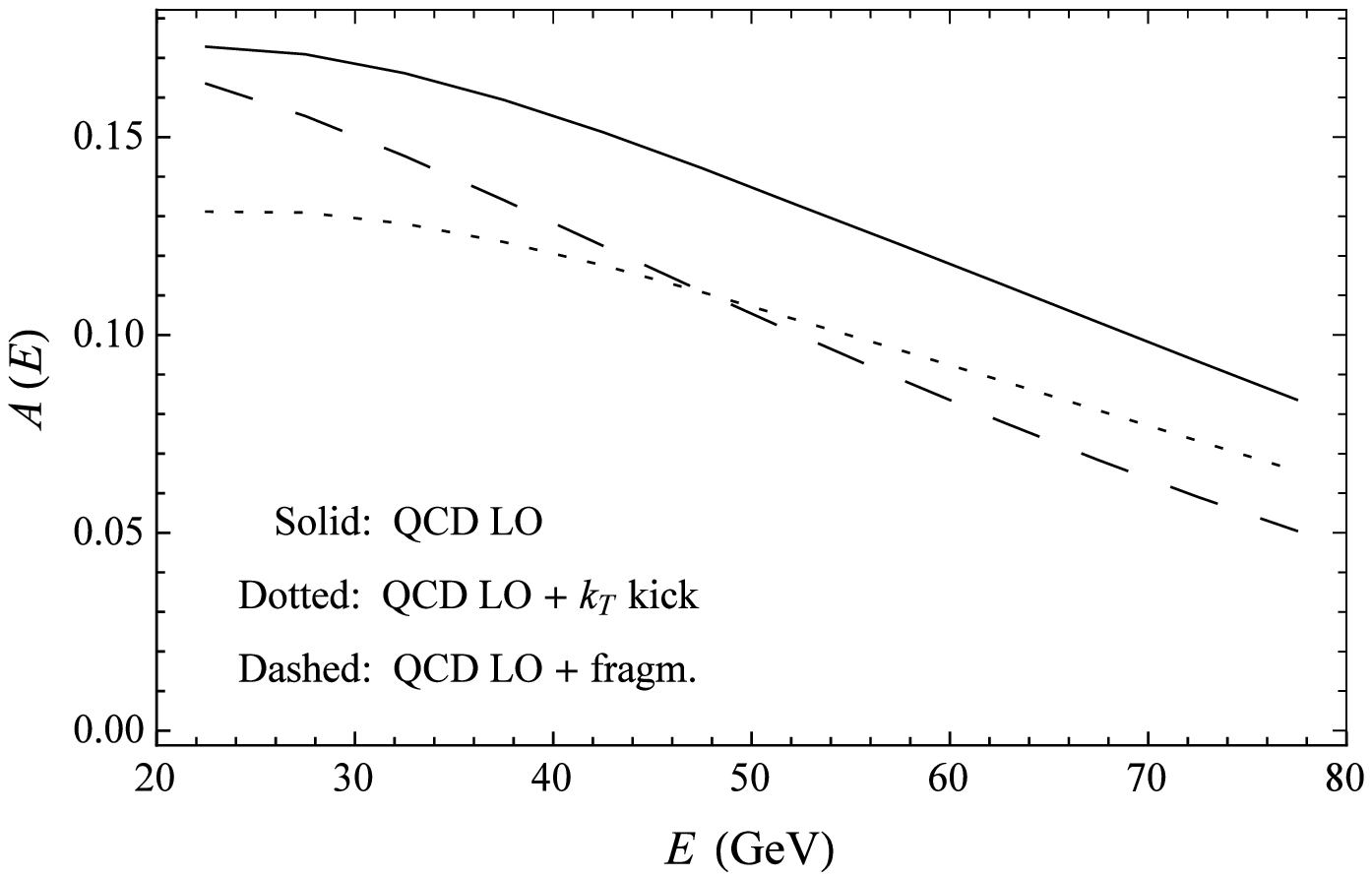,width=220pt}}
& \mbox{\epsfig{file=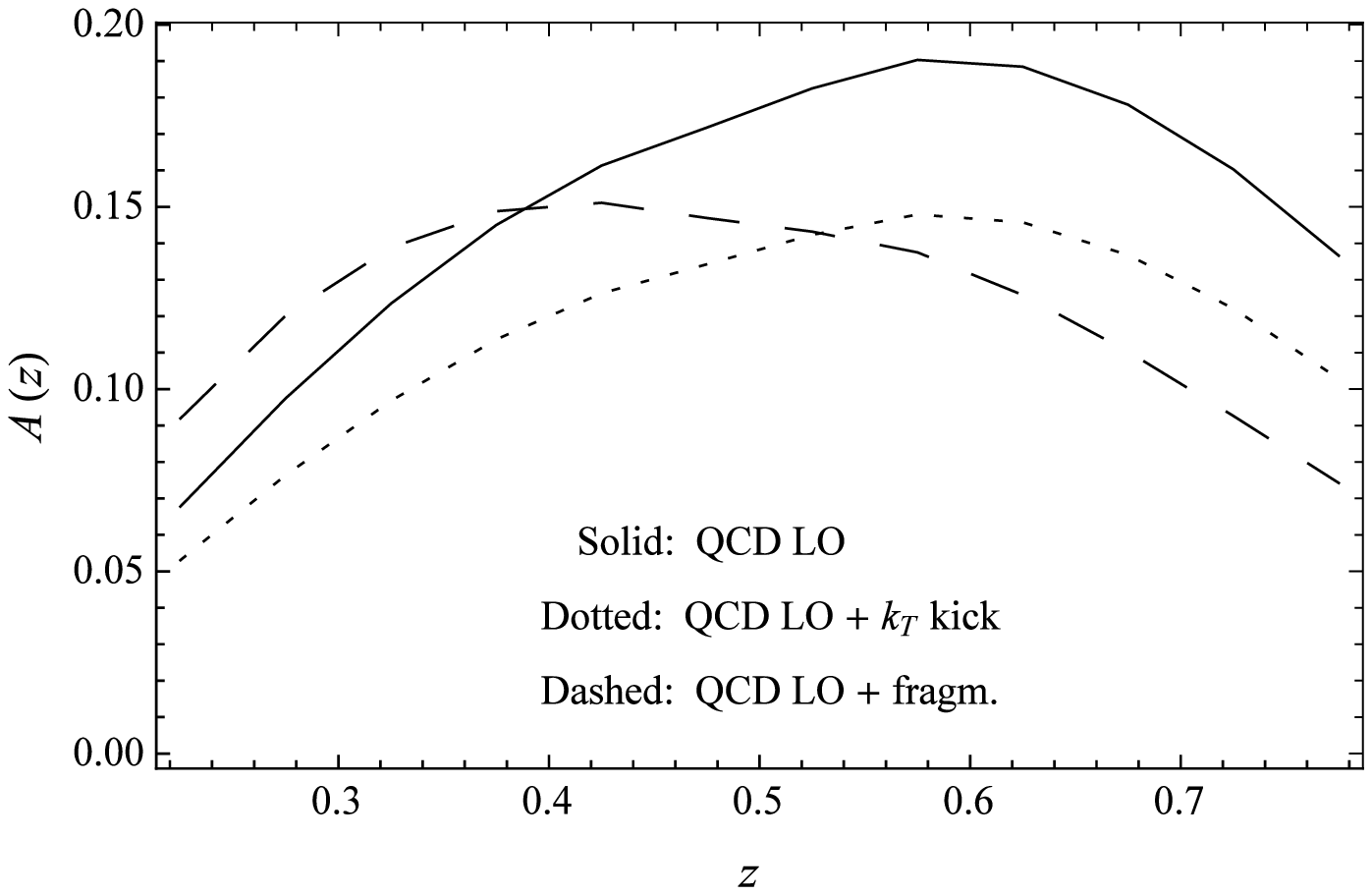,width=220pt}}\\
\mbox{\epsfig{file=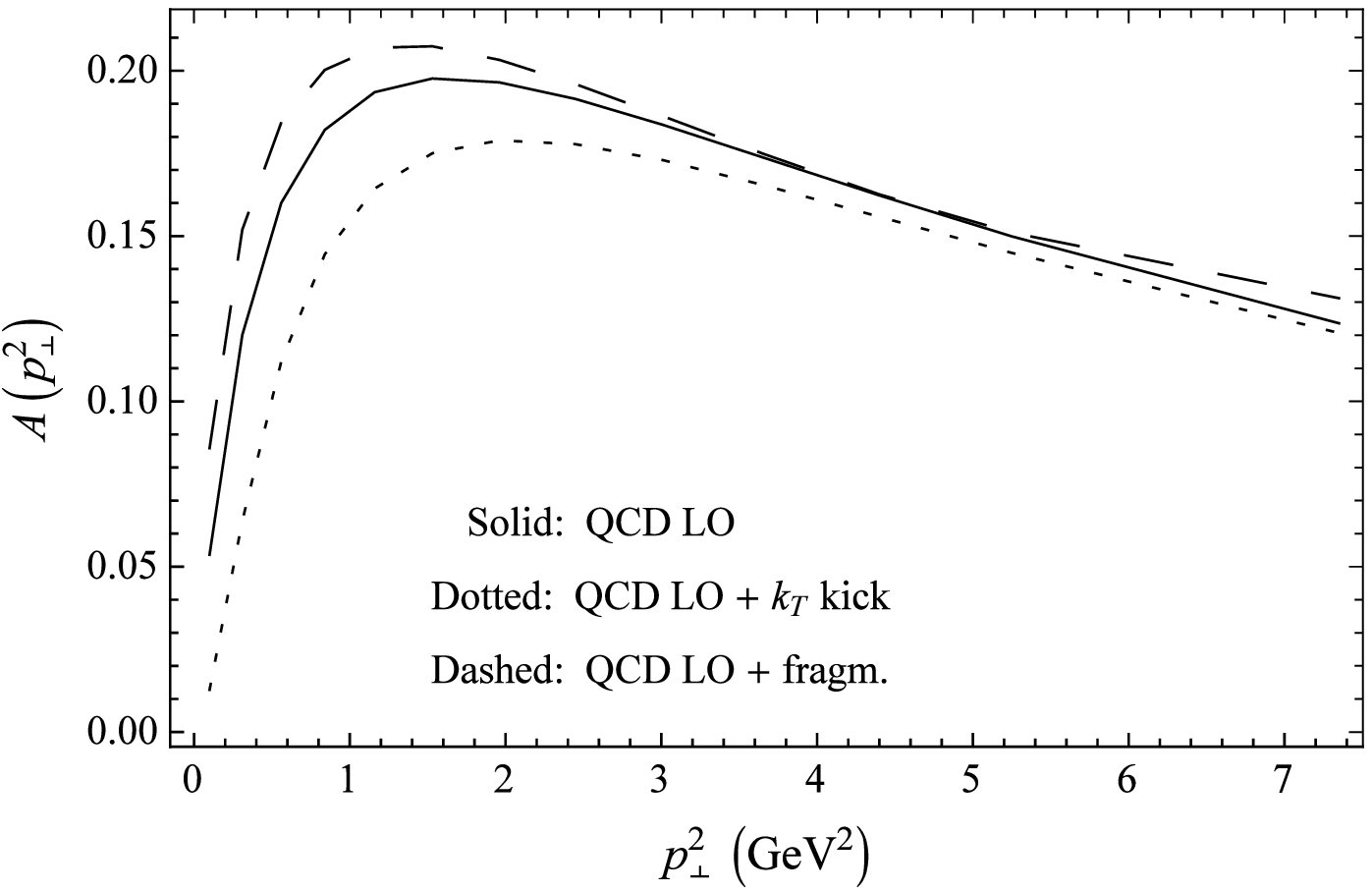,width=220pt}}
& \mbox{\epsfig{file=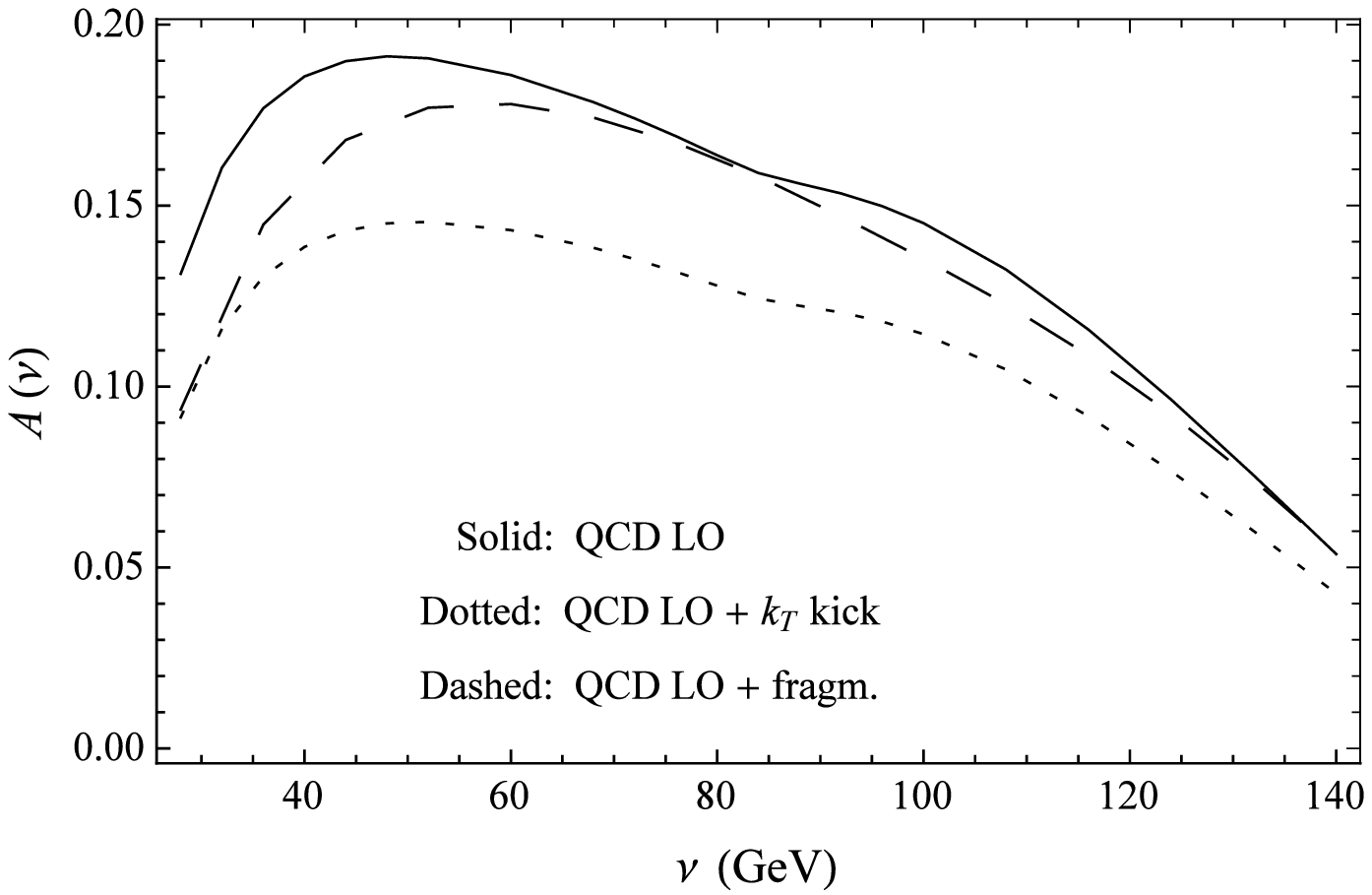,width=220pt}}\\
\end{tabular}
\caption{\label{Fg.3}\small Semi-inclusive differential distributions of the $\cos2\varphi$ asymmetry for the COMPASS kinematics as a function of the charmed meson lab energy $E$, fractional energy $z=\frac{E}{\nu}$, squared transverse momentum $p^{2}_{\perp}$ and virtual photon energy $\nu$. For all distributions, the LO (solid curves), LO + $k_{T}$-kick (dotted curves) and LO + fragmentation (dashed curves) predictions are shown.}
\end{center}
\end{figure}

Our results for differential distributions of the AA in charm leptoproduction are presented in Fig.~\ref{Fg.3}. We consider the $E$-, $z$-, $p^{2}_{\perp}$- and $\nu$-distributions of the asymmetry defined as 
\begin{align}
A(E)&=\frac{2\int\limits_{0}^{2\pi }{\rm d}\varphi \cos 2\varphi \dfrac{{\rm d}^{2}\sigma _{lN}}{{\rm d}E{\rm d}\varphi}(E,\varphi)}{\int\limits_{0}^{2\pi }{\rm d}\varphi \dfrac{{\rm d}^{2}\sigma _{lN}}{{\rm d}E{\rm d}\varphi}(E,\varphi)},& A(z)&=\frac{2\int\limits_{0}^{2\pi }{\rm d}\varphi \cos 2\varphi \dfrac{{\rm d}^{2}\sigma _{lN}}{{\rm d}z\,{\rm d}\varphi}(z,\varphi)}{\int\limits_{0}^{2\pi }{\rm d}\varphi \dfrac{{\rm d}^{2}\sigma _{lN}}{{\rm d}z\,{\rm d}\varphi}(z,\varphi)}, \label{16} \\
A(p^{2}_{\perp})&=\frac{2\int\limits_{0}^{2\pi }{\rm d}\varphi \cos 2\varphi \dfrac{{\rm d}^{2}\sigma _{lN}}{{\rm d}p^{2}_{\perp}{\rm d}\varphi}(p^{2}_{\perp},\varphi)}{\int\limits_{0}^{2\pi }{\rm d}\varphi \dfrac{{\rm d}^{2}\sigma _{lN}}{{\rm d}p^{2}_{\perp}{\rm d}\varphi}(p^{2}_{\perp},\varphi)},& A(\nu)&=\frac{2\int\limits_{0}^{2\pi }{\rm d}\varphi \cos 2\varphi \dfrac{{\rm d}^{2}\sigma _{lN}}{{\rm d}\nu{\rm d}\varphi}(\nu,\varphi)}{\int\limits_{0}^{2\pi }{\rm d}\varphi \dfrac{{\rm d}^{2}\sigma _{lN}}{{\rm d}\nu{\rm d}\varphi}(\nu,\varphi)}. \notag 
\end{align}

Note that all the computations are performed for the kinematics of the COMPASS experiment at CERN, i.e., for $E_{l}$=160 GeV, 0.003 GeV$^2$~$< Q^2 <$~10 GeV$^2$, $3\cdot 10^{-5} < x < 0.1$ and 20~GeV $< E <$~80 GeV. 

In our calculations, the CT14nlo parametrization of the PDF together with the value $m_c=1.25$~GeV \cite{CT14} are used. Unless otherwise stated, we use $\mu_{F}=\sqrt{4m_c^{2}+Q^{2}}$ throughout this paper.

One of the most remarkable properties of the asymmetry is its parametric stability. Our analysis shows that the fixed-order predictions for the AA are less sensitive to standard uncertainties in the QCD input parameters than the corresponding ones for the production
cross sections. For instance, sufficiently above the production threshold, changes of $\mu_F$ in the range $(1/2)\sqrt{4m_{c}^{2}+Q^{2}}<\mu <2\sqrt{4m_{c}^{2}+Q^{2}}$ only lead to a few percent variations of $A(E)$. We analyze also the dependence of the pQCD
predictions on the uncertainties in the heavy-quark mass. We observe that changes of the $c$-quark mass in the interval 1.3 $<m_{c}<1.7$ GeV affect the AA by about (5--7)\% at $Q^{2}<10$ GeV$^2$ and $x<10^{-1}$. The corresponding variations of the cross sections are larger by an order of magnitude. We also verified that the pQCD predictions for the AA are practically independent of the gluon distribution function, $g(\zeta,\mu_F)$, in use.

One can see from Fig.~\ref{Fg.3} that the $c$-quark fragmentation has different impact on the   
$E$-, $z$-, $p^{2}_{\perp}$-, and $\nu$-distributions of the azimuthal $\cos2\varphi$  asymmetry. Since the azimuth-dependent and independent cross sections have different energy behavior, the convolution (\ref{15}) affects essentially the Born-level predictions for $A(E)$, especially at high values of $E$. The same situation takes also place for $A(z)$ where $z=\frac{E}{\nu}$ is the fractional energy. Were the integration done over entire range of the $D$-meson energy $E$, the QCD predictions for $A(\nu)$ with and without fragmentation  would be exactly equivalent. This is because the variable $\nu=\frac{y\bar{S}-m_{N}^{2}}{2m_{N}}$ is independent of the $c$-quark 4-momentum, $p_{Q}$. However, the COMPASS spectrometer has the acceptance 20~GeV $< E <$~80 GeV that leads to small variations of the quantity $A(\nu)$ under the fragmentation convolution (\ref{15}). For the $A(p^{2}_{\perp})$ distribution, the hadronization effects are also expected to be unessential. 
 
One can see from Fig.~\ref{Fg.3} that the $k_{T}$-kick contribution (dotted curve) to the $A(E)$ spectrum is sizable in the region $E <$~40 GeV. At higher  energies, the hadronization mechanism (dashed curve) becomes essential. The $A(\nu)$ spectrum is stable under the $c$-quark fragmentation but sensitive to the gluon transverse motion in the target. Since the AA is stable under radiative corrections, we expect that measurement of its   differential distributions  in wide kinematic range will improve our knowledge of both the gluon TMD distribution and charm fragmentation function.

As to the azimuthal $\cos\varphi$ asymmetry in heavy flavor production, it is expected to be small. Were the integration done over entire range of the $D$-meson energy $E$, the LO QCD predictions for the $\cos\varphi$ asymmetry would be vanishing because the corresponding partonic cross section, $B_{I}(\hat{x},Q^{2},t_{1},u_{1})$, in Eq.(\ref{9}) is anti-symmetric under $t_{1}\leftrightarrow u_{1}$. We have verified numerically that the quantity
\begin{equation}  \label{17}
A^{\cos\varphi}(E)=\frac{2\int{\rm d}\varphi \cos\varphi\,{\rm d}^{2}\sigma _{lN}(E,\varphi)}{\int{\rm d}\varphi\,{\rm d}^{2}\sigma _{lN}(E,\varphi)}
\end{equation}
is of the order of 1\% in the COMPASS kinematics.

\subsection{Production cross sections}
\label{3.2}

\begin{figure}[t]
\begin{center}
\begin{tabular}{cc}
\mbox{\epsfig{file=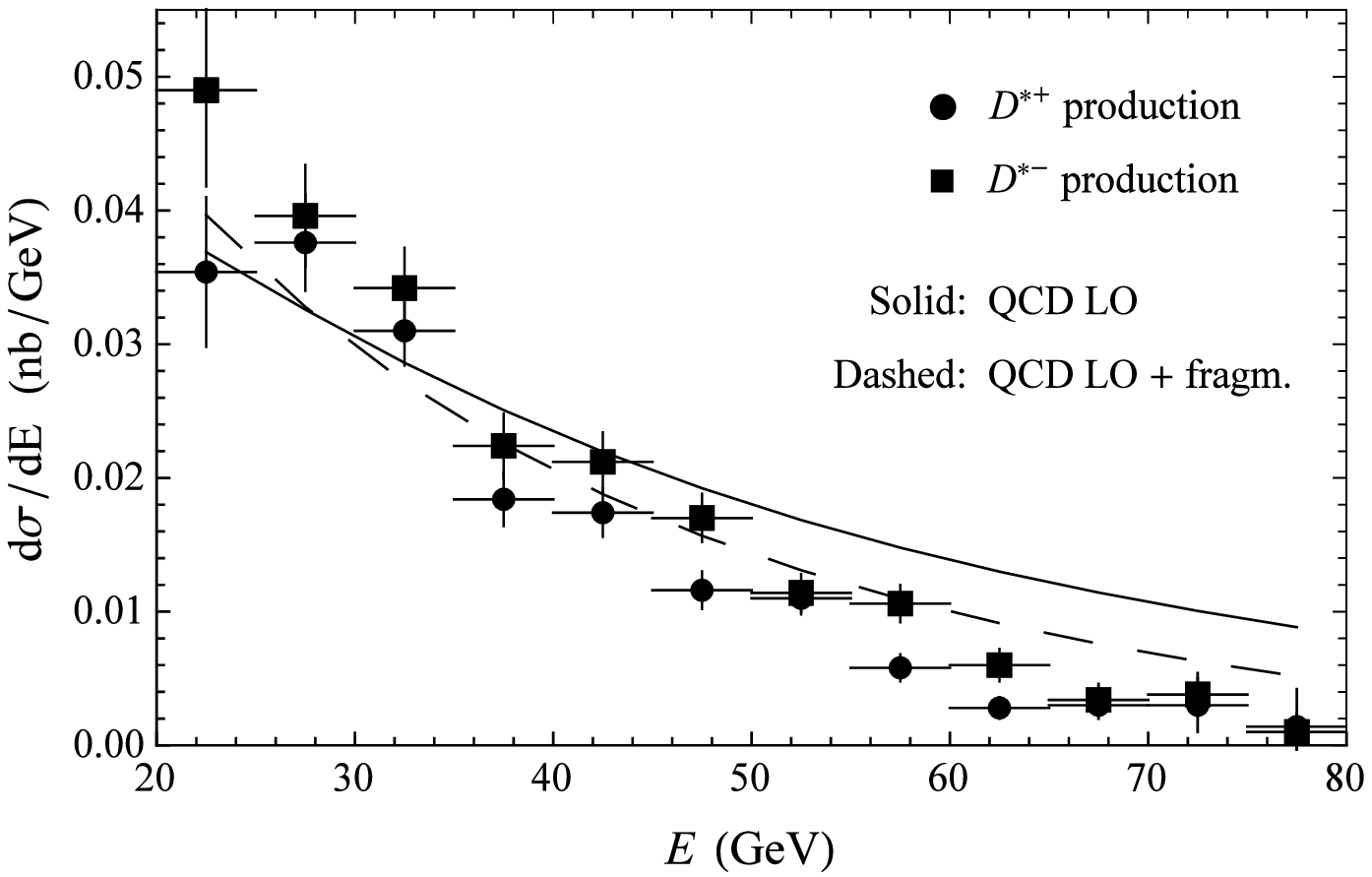,width=220pt}}
& \mbox{\epsfig{file=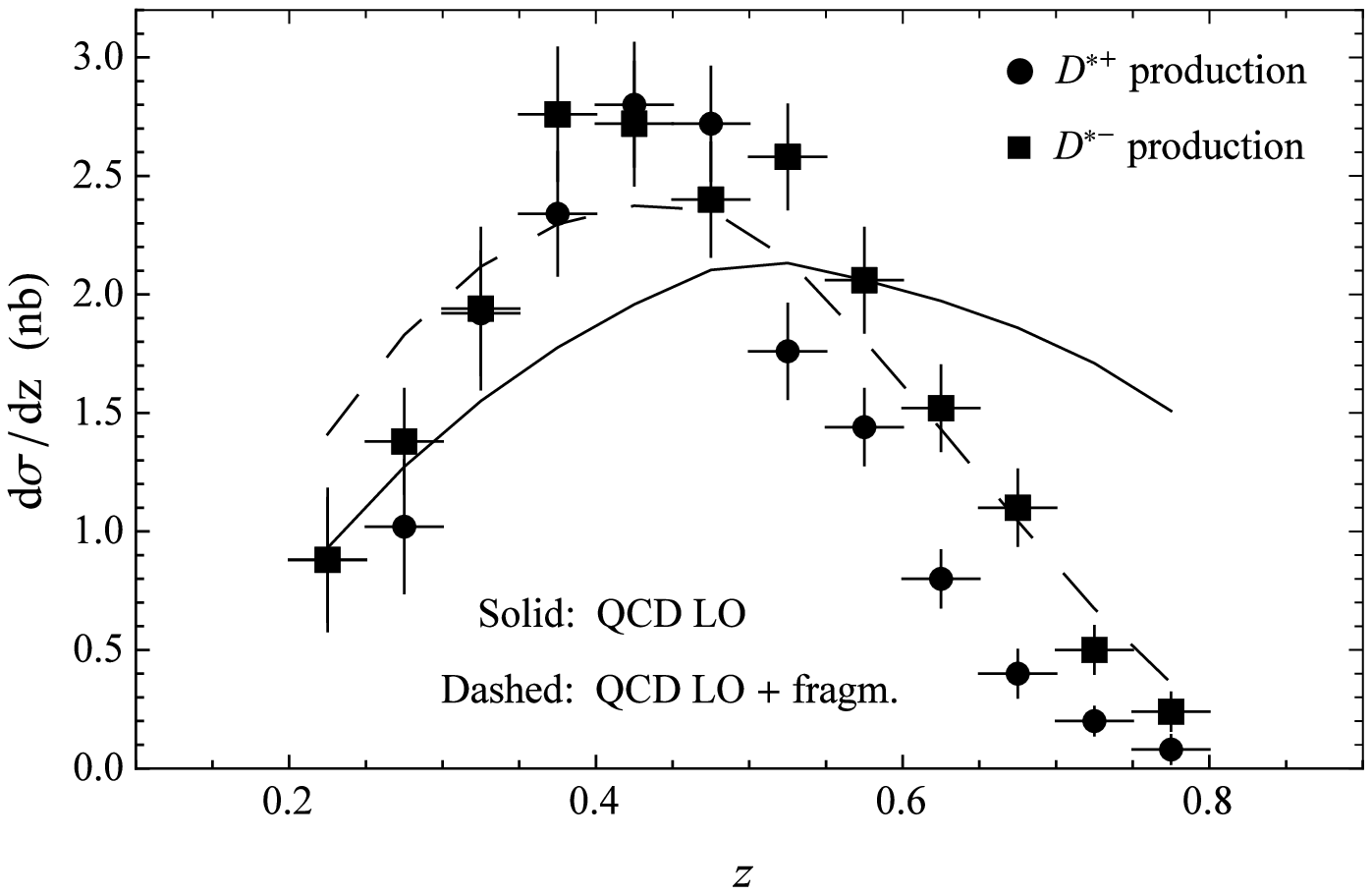,width=220pt}}\\
\mbox{\epsfig{file=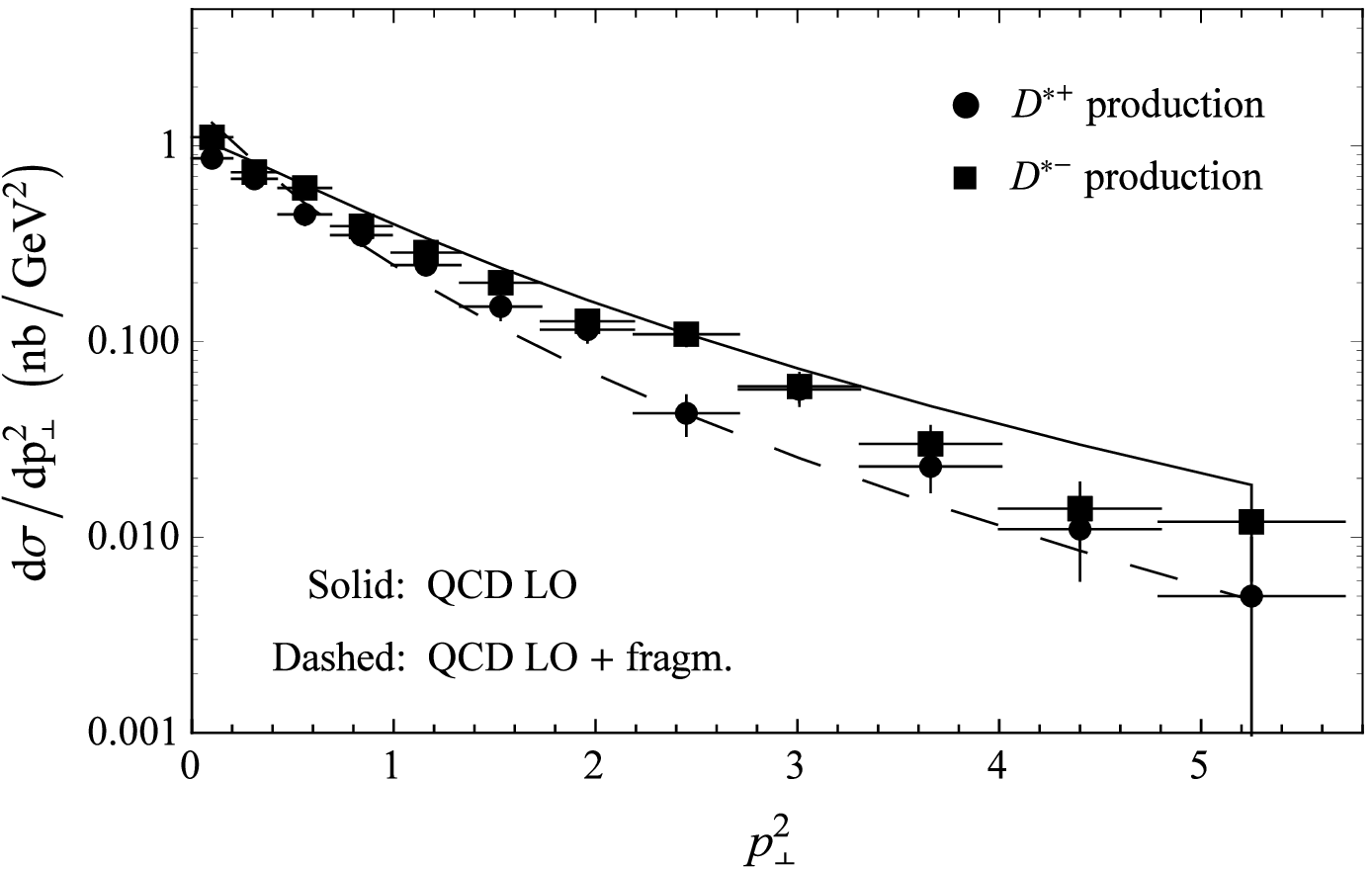,width=220pt}}
& \mbox{\epsfig{file=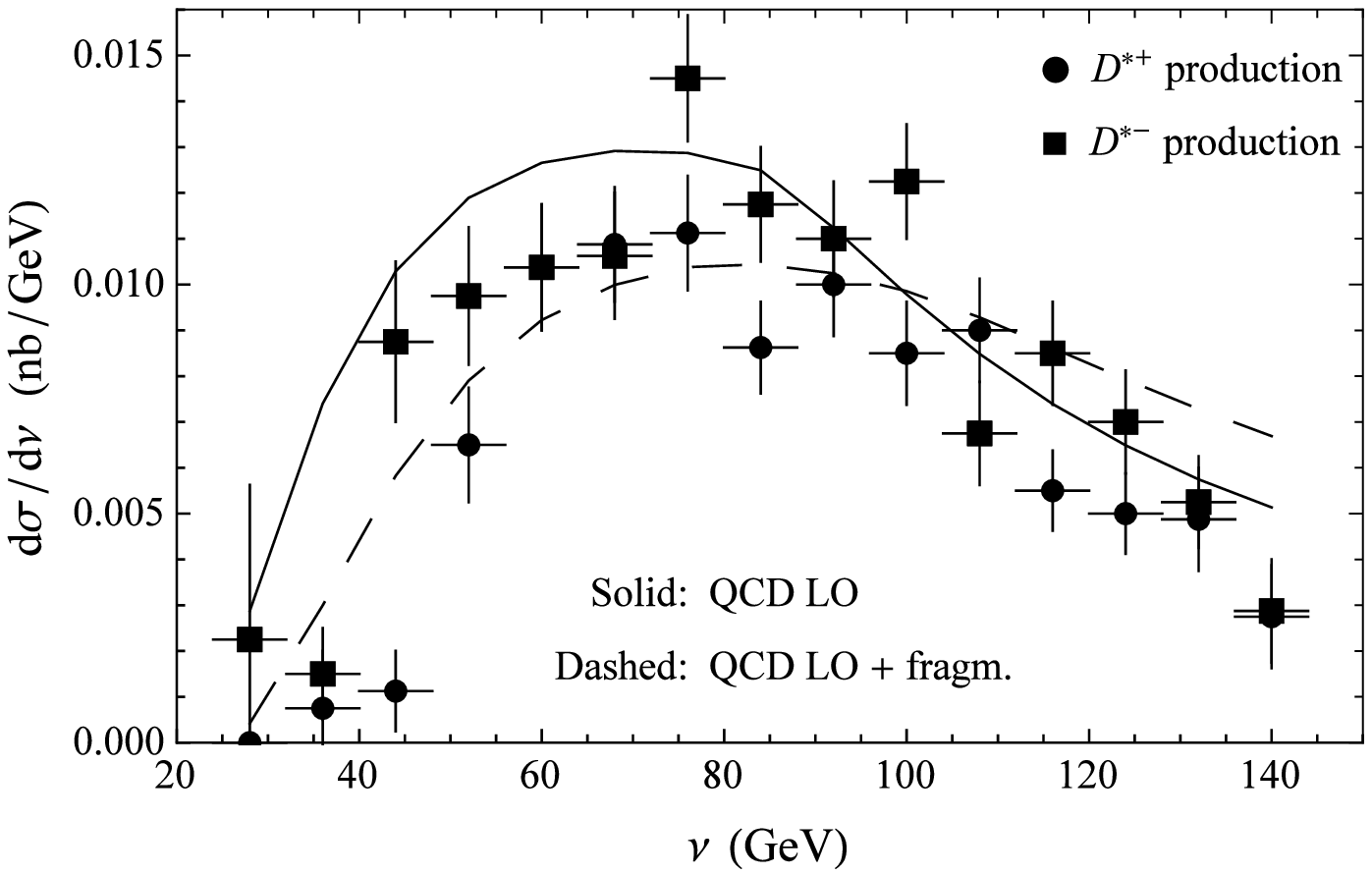,width=220pt}}\\
\end{tabular}
\caption{\label{Fg.4}\small Semi-inclusive differential cross sections for $D^{*\pm}$ mesons production as a function of the charmed meson lab energy $E$, fractional energy $z=\frac{E}{\nu}$, squared transverse momentum $p^{2}_{\perp}$ and virtual photon energy $\nu$. Plotted are the COMPASS data \cite{Compass_2012} for $D^{*+}$ (circles)  and $D^{*-}$ (squares) mesons, as well as the QCD LO (solid curves)  and LO + fragmentation (dashed curves) predictions.}
\end{center}
\end{figure}

In Fig.~\ref{Fg.4}, we compare the COMPASS data \cite{Compass_2012} on the semi-inclusive differential cross sections for $D^{*+}$ and $D^{*-}$ mesons production (black circles and squares, respectively) with the QCD LO and LO + fragmentation predictions (solid and dashed lines, correspondingly). Note that, in the considered approximation (i.e., in the LO photon-gluon fusion), there is no any  difference between $D^{*+}$ and $D^{*-}$ mesons distributions.\footnote{To describe the difference between $D^{*+}$ and $D^{*-}$ mesons spectra, one should take into account the light quark contributions.} To normalize the fragmentation function, we use the common assumption of 0.6 $D^{*}$ mesons per charm event. One can see from Fig.~\ref{Fg.4} that predictions of the considered simplified approach (i.e., QCD LO + Peterson fragmentation) are in a reasonable agreement with the experimental results \cite{Compass_2012} on the $D^{*}$ mesons production cross sections.

\section{Conclusion}
In this paper, we study the QCD predictions for the azimuthal $\cos 2\varphi$ asymmetry in charm leptoproduction for the kinematics of the COMPASS spectrometer. Since the next-to-leading order corrections to the AA are predicted to be small, we restrict ourselves by the leading order consideration and concentrate on the nonperturbative contributions: the gluon transverse motion in the target and $c$-quark fragmentation. Because of the $c$-quark low mass, these nonperturbative contributions to the AA are predicted to be sizable, about (30--40)\%. As to the $\cos \varphi$ asymmetry in charm leptoproduction, it is predicted to be small  (about 1\%) in the considered kinematics. We conclude that extraction of the azimuthal asymmetries from available COMPASS data will provide valuable information about the TMD distribution of the gluon in the proton and the $c$-quark hadronization mechanism.

Presently, the AAs in heavy-quark electroproduction are completely unmeasured. Experimental information about the $\cos 2\varphi$ asymmetry could justify its (predicted) remarkable properties and provide very promising applications. In particular, the azimuthal asymmetry seems to be good probe of the intrinsic charm density \cite{we5,we6} and linearly polarized gluon distribution, $h_{1}^{\perp g}\big(\zeta,\vec{k}_{T}^2\big)$, in unpolarized proton.\footnote{The circular gluon polarization, $\langle\Delta g/ g \rangle$, inside longitudinally polarized nucleons has been measured by the COMPASS Collaboration using the double spin asymmetry in open charm muoproduction  \cite{Compass_2009}. Note also the paper \cite{Mukherjee_2016}, where a possibility to probe the function $h_{1}^{\perp g}\big(\zeta,\vec{k}_{T}^2\big)$ using the $J/\psi$ production in unpolarized $pp$ collisions was considered.} (For details, see Appendix).

\section*{Acknowledgements} The authors are grateful to S.~J.~Brodsky, A.~V.~Kotikov, O.~Kouznetsov, A.~B.~Kniehl, E.~Leader, S.~O.~Moch and A.~G.~Oganesian for useful discussions. This work is supported in part by the State Committee of Science of RA, grant 15T-1C223. N.Ya.I. thanks BLTP for kind hospitality and Ter-Antonyan - Smorodinsky program for support of his visit to JINR. 

\appendix

\section{Azimuthal asymmetries and linear polarization of gluons inside unpolarized nucleon}
Let us now discuss how the azimuthal asymmetries under consideration can be used in extraction of the gluonic analogue of the Boer-Mulders function, $h_{1}^{\perp g}\big(\zeta,\vec{k}_{T}^2\big)$, from future data on heavy-quark pair production in lepton-nucleon DIS. The function $h_{1}^{\perp g}\big(\zeta,\vec{k}_{T}^2\big)$ describes the distribution of linearly polarized gluons inside unpolarized nucleon \cite{Mulders_2001}. The corresponding TMD distribution of unpolarized gluons is usually denoted by $f_{1}^{g}\big(\zeta,\vec{k}_{T}^2\big)$,\, $g(\zeta)=\int {\rm d}^2 k_{T} f_{1}^{g}\big(\zeta,\vec{k}_{T}^2\big)$. In Refs.\cite{Boer_HQ_1,Boer_HQ_2,Boer_HQ_3}, it was proposed to study the linearly polarized gluon density in unpolarized nucleon using the heavy-quark pair production in the reaction
\begin{equation} \label{18}
l(\ell )+N(P)\rightarrow l^{\prime}(\ell -q)+Q(p_{Q})+\bar{Q}(p_{\bar{Q}})+X(p_{X}). 
\end{equation}
For this purpose, the momenta of both heavy quark and anti-quark, $\vec{p}_{Q}$ and $\vec{p}_{\bar{Q}}$, in the process (\ref{18}) should be measured (reconstructed). For further analysis, the sum and difference of the transverse heavy quark momenta are introduced,
\begin{align} \label{19}
\vec{K}_{\perp}&=\frac{1}{2}\left(\vec{p}_{Q\perp}-\vec{p}_{\bar{Q}\perp}\right), &
\vec{k}_{T}&=\vec{p}_{Q\perp}+\vec{p}_{\bar{Q}\perp},
\end{align}
in the plane orthogonal to the direction of the target and the exchanged photon. The azimuthal angles of $\vec{K}_{\perp}$ and $\vec{k}_{T}$ (relative to the the lepton scattering plane projection, $\phi_l=\phi_{l^{\prime}}=0$) are denoted by $\phi_{\perp}$ and $\phi_{T}$,  respectively. It is also useful to introduce the sum and difference of the magnitudes of the heavy quark transverse momenta,
\begin{align} 
K&=\frac{1}{2}\left(\left|\vec{p}_{Q\perp}\right|+\left|\vec{p}_{\bar{Q}\perp}\right|\right), & \vec{K}_{\perp}^2&=\frac{1}{4}\left(\Delta K\right)^2\sin^2\frac{\alpha}{2}+K^2\cos^2\frac{\alpha}{2}, \notag  \\
\Delta K&=\left|\vec{p}_{Q\perp}\right|-\left|\vec{p}_{\bar{Q}\perp}\right|, & \vec{k}_{T}^2&=\left(\Delta K\right)^2\cos^2\frac{\alpha}{2}+4K^2\sin^2\frac{\alpha}{2}, \label{20}
\end{align}
where $\alpha=\pi-(\varphi_{Q}-\varphi_{\bar{Q}})$ and $\varphi_{Q}$ ($\varphi_{\bar{Q}}$) is the azimuth of the detected quark (anti-quark).

At LO, ${\cal O}(\alpha _{em}\alpha_{s})$, the only parton-level subprocess for the reaction (\ref{18}) is $\gamma^{*}(q)+g(k_g)\rightarrow Q(p_{Q})+\bar{Q}(p_{\bar{Q}})$, where $\vec{k}_{g}\simeq \zeta\vec{p}+\vec{k}_{T}$. For this reason, the quantity $\Delta K$ is determined by the gluon transverse momentum in the target, $\Delta K\leq \big|\,\vec{k}_{T}\big|\sim \Lambda_{{\rm QCD}}$. Sizable values for the azimuthal asymmetries are expected at $K\sim \left|\vec{p}_{Q\perp}\right|\gtrsim m$. In this kinematics (i.e., for $\Delta K\big/K\sim \Lambda_{{\rm QCD}}\big/\,m\ll 1$), the following relations between the azimuthal angles hold:
\begin{align}
\phi_{\perp}&\simeq \frac{\varphi_{Q}+\varphi_{\bar{Q}}}{2}+\frac{\pi}{2}=\varphi_{Q}+ \frac{\alpha}{2},& \alpha &=\pi-(\varphi_{Q}-\varphi_{\bar{Q}}), \notag \\
\phi_{T}&\simeq \frac{\varphi_{Q}+\varphi_{\bar{Q}}}{2}=\varphi_{Q}+\frac{\alpha-\pi}{2},& \phi_{T}&\simeq \phi_{\perp}-\frac{\pi}{2}.  \label{21} 
\end{align}
One can see from Eqs.(\ref{21}) that the angles $\phi_{\perp}$ and $\phi_{T}$ are not independent from each other at $\Delta K\big/K\ll 1$. 
In this approximation, the master formula (21) in Ref.~\cite{Boer_HQ_2} for the angular structure of the cross section (\ref{18}) takes the following form:
\begin{equation} \label{22}
{\rm d}\sigma_{lN}\propto \left\{ A_0- \vec{k}^{2}_{T}B_0 + \left[A_1- \vec{k}^{2}_{T} \left(B_1+B_1^{\prime}\right)\right]\cos\phi_{\perp}+ \left[A_2-\vec{k}^{2}_{T}\left(B_2+B_2^{\prime}\right)\right]\cos2\phi_{\perp}\right\}.
\end{equation}
Corrections to the approximate Eqs.(\ref{21},\ref{22}) are of the order of ${\cal O}(\Delta K\big/K)$. Note also that $\Delta K\big/K\ll 1$ implies $\big|\,\vec{k}_{T}\big|\ll \big|\vec{K}_{\perp}\big|$ for $\varphi_{Q}-\varphi_{\bar{Q}}\approx \pi$.

The exact definitions and LO predictions for the terms $A_i$, $B_i$ ($i=0,1,2$) and $B_{1,2}^{\prime}$  are presented in Refs.~\cite{Boer_HQ_2,Boer_HQ_3}. The quantities $A_i$ are determined by the unpolarized TMD gluon distribution, $A_i=\hat{A}_i\, f_{1}^{g}\big(\zeta^{-},\vec{k}_{T}^2\big)$, while $B_i$  and $B_{i}^{\prime}$ depend on the linearly polarized gluon density, $B_i^{(\prime)}=\frac{1}{2 m^2_N}\hat{B}_i^{(\prime)}\,h_{1}^{\perp g}\big(\zeta^{-},\vec{k}_{T}^2\big)$, where $\zeta^{-}=\frac{-U_1}{y\bar{S}+T_1}$. 

Integrating Eq.(\ref{22}) over the anti-quark azimuth, $\varphi_{\bar{Q}}$, we obtain the following angular structure at $\big|\,\vec{k}_{T}\big|\ll \big|\vec{K}_{\perp}\big|$:
\begin{equation} \label{23}
\frac{{\rm d}^{6}\sigma_{lN}}{{\rm d}x\,{\rm d}Q^{2}{\rm d}T_{1}{\rm d}\vec{K}^2_{\perp}{\rm d}\vec{k}^2_{T}{\rm d}\varphi}= {\cal N} A_0\left\{1 - \vec{k}^{2}_{T}\frac{B_0}{A_0} + \frac{A_1}{A_0}\left[1- \vec{k}^{2}_{T}\frac{B_1+B_1^{\prime}}{A_1}\right]\cos\varphi+ \frac{A_2}{A_0}\left[1- \vec{k}^{2}_{T}\frac{B_2+B_2^{\prime}}{A_2}\right]\cos2\varphi\right\},
\end{equation}
where ${\cal N}$ is a normalization factor, while $\varphi\equiv \varphi_{Q}$ and $\vec{K}^2_{\perp}=p^2_{\perp}$ are the heavy-quark azimuth and transverse momentum  defined by Eqs.(\ref{4}) and (\ref{5}), respectively.

Eq.(\ref{23}) is obtained within the TMD factorization scheme and can serve as a generalization of the cross section (\ref{2}) which was derived in the framework of collinear factorization. Putting $h_{1}^{\perp g}\big(\zeta,\vec{k}_{T}^2\big)=\delta\big(\vec{k}_{T}^2\big)\,\hat{h}_{1}^{\perp g}(\zeta)$ and integrating Eq.(\ref{23}) over $\vec{k}_{T}^2$, we will exactly reproduce the collinear results given by Eqs.(\ref{2}-\ref{9}).
 
One can see from Eq.(\ref{23}) that the gluonic analogue of the Boer-Mulders function, $h_{1}^{\perp g}\big(\zeta,\vec{k}_{T}^2\big)$, can, in principle, be determined from measurements of the $\vec{k}^{2}_{T}$-dependence of the $\cos\varphi$ and $\cos 2\varphi$ asymmetries.  Within the TMD scheme, these asymmetries have the form $\frac{A_1}{A_0}\left(1- \vec{k}^{2}_{T}\frac{B_1+B_1^{\prime}}{A_1}\right)$ and $\frac{A_2}{A_0}\left(1- \vec{k}^{2}_{T}\frac{B_2+B_2^{\prime}}{A_2}\right)$, respectively. The ratios $\frac{A_1}{A_0}$ and $\frac{A_2}{A_0}$ describe the $\cos\varphi$ and $\cos 2\varphi$ distributions in the collinear scheme, i.e., $\frac{A_1}{A_0}\approx \frac{(2-y)\sqrt{1-y}}{1+(1-y)^2} \frac{{\rm d}^2\sigma_{I}}{{\rm d}^2\sigma_{2}}$ and $\frac{A_2}{A_0}\approx \frac{2(1-y)}{1+(1-y)^2} \frac{{\rm d}^2\sigma_{A}}{{\rm d}^2\sigma_{2}}$, where the cross sections ${\rm d}^2\sigma_{k} $ ($k=2,L,A,I$) are defined by Eqs.(\ref{2}-\ref{9}).

As noted in Section \ref{3.0}, the quantity $\frac{A_1}{A_0}\sim \frac{{\rm d}^2\sigma_{I}}{{\rm d}^2\sigma_{2}}$ is predicted to be small, of the order of 1\%. For this reason, it will be questionable to extract experimentally the ratio $\vec{k}^{2}_{T}\frac{B_1+B_1^{\prime}}{A_1}= \frac{\vec{k}^{2}_{T}}{2m_N^2}\frac{h_{1}^{\perp g}\big(\zeta,\vec{k}_{T}^2\big)}{f_{1}^{g}\big(\zeta,\vec{k}_{T}^2\big)}$ from future measurements of the $\cos \varphi$ distribution.

At the same time, the ratio $\frac{A_2}{A_0}\sim \frac{{\rm d}^2\sigma_{A}}{{\rm d}^2\sigma_{2}}$ is predicted to be large, about (15--20)\%, and perturbatively stable. We conclude that the $\cos 2\varphi$ asymmetry in heavy-quark leptoproduction could be good  probe of the linear polarization of gluons inside unpolarized nucleon. Concerning the experimental aspects, azimuthal asymmetries in charm and bottom production can be measured at the proposed EIC \cite{EIC} and LHeC \cite{LHeC2} colliders. In detail, our predictions for the $\cos 2\varphi$ asymmetry within the TMD factorization scheme will be presented in a forthcoming publication.

\vspace{3mm}



\end{document}